\documentclass[%
 aip,
 jmp,%
 amsmath,amssymb,
reprint,%
]{revtex4-1}

\usepackage{graphicx}
\usepackage{dcolumn}
\usepackage{bm}
\usepackage{braket}
\usepackage{bbold}

\begin{document}

\preprint{APS/123-QED}

\title{Path to perfect photon entanglement with a quantum dot}
%
 \author{A. Fognini}
 \email{a.w.fognini@tudelft.nl}
 \affiliation{
 Kavli Institute of Nanoscience Delft,\\
 Delft University of Technology, Delft 2628 CJ, The Netherlands
 }%
 \author{A. Ahmadi}
  \email{arash.ahmadi@uwaterloo.ca}
  \affiliation{%
   Institute for Quantum Computing and Department of Physics \& Astronomy, \\
   University of Waterloo, Waterloo, ON N2L 3G1, Canada
 }%
 
 \author{M. Zeeshan}
  \affiliation{%
   Institute for Quantum Computing and Department of Electrical \& Computer Engineering, \\
   University of Waterloo, Waterloo, ON N2L 3G1, Canada
 }%
 
 \author{J. T. Fokkens}
 \affiliation{
 Kavli Institute of Nanoscience Delft,\\
 Delft University of Technology, Delft 2628 CJ, The Netherlands 
 }%
 \author{S. J. Gibson}
  \affiliation{%
   Institute for Quantum Computing and Department of Electrical \& Computer Engineering, \\
   University of Waterloo, Waterloo, ON N2L 3G1, Canada
 }%
 \author{N. Sherlekar}
  \affiliation{%
   Institute for Quantum Computing and Department of Electrical \& Computer Engineering, \\
   University of Waterloo, Waterloo, ON N2L 3G1, Canada
 }%
 \author{S. J. Daley}
  \affiliation{%
   Institute for Quantum Computing and Department of Electrical \& Computer Engineering, \\
   University of Waterloo, Waterloo, ON N2L 3G1, Canada
 }%
 \author{D. Dalacu}%
  \affiliation{%
  National Research Council of Canada, Ottawa, ON K1A 0R6, Canada
 }%

  \author{P. J. Poole}%
  \affiliation{%
  National Research Council of Canada, Ottawa, ON K1A 0R6, Canada
 }%
 \author{K.~D.~J\"ons}%
  \affiliation{%
   Department of Applied Physics, Royal Institute of Technology (KTH), \\
   AlbaNova University Center, SE - 106 91 Stockholm, Sweden
 }%
 \author{V. Zwiller}%
  \affiliation{%
   Department of Applied Physics, Royal Institute of Technology (KTH), \\
   AlbaNova University Center, SE - 106 91 Stockholm, Sweden
 }%
 \affiliation{
 Kavli Institute of Nanoscience Delft,\\
 Delft University of Technology, Delft 2628 CJ, The Netherlands
 }%
 \author{M. E. Reimer}
  \affiliation{%
   Institute for Quantum Computing and Department of Electrical \& Computer Engineering, \\
   University of Waterloo, Waterloo, ON N2L 3G1, Canada
 }

%

\date{\today}

\begin{abstract}
Realizing perfect two-photon entanglement from quantum dots has been a long-standing scientific challenge. It is generally thought that the nuclear spins limit the entanglement fidelity through spin flip dephasing processes. However, this assumption lacks experimental support. Here, we show dephasing-free two-photon entanglement from an Indium rich single quantum dot comprising of nuclear spin 9/2 when excited quasi-resonantly. This remarkable finding is based on a perfect match between our entanglement measurements with our model that assumes no dephasing and takes into account the detection system's timing jitter and dark counts. We discover that neglecting the detection system is responsible for not reaching perfect entanglement in the past and not the nuclear spins. Therefore, the key to unity entanglement from quantum dots comprises of a resonant excitation scheme and a detection system with ultra-low timing jitter and dark counts.

\end{abstract}

\maketitle







Quantum dots can generate polarization entangled photons through the biexciton-exciton cascade \cite{PhysRevLett.84.2513, hafenbrak, salterNat}. Understanding how this process can yield perfect polarization entanglement has been a significant scientific challenge for more than a decade. Still, the experimental demonstration of perfect entanglement from quantum dots (QDs) remains elusive \cite{Huber2017, Keil2017}. The reason is twofold. First, QDs must emit perfectly entangled photons, and second, the detection system must be capable of measuring it without degrading its value \cite{PhysRevLett.101.170501}. Up to now, the detrimental effects of the detection system have been mostly ignored. Nonetheless, they are of equal importance to the photon generation process as timing jitter and dark counts can spoil the measured entanglement. Here, we show that it is possible to reach dephasing free entanglement from QDs by considering both the generation and detection processes of the entangled photons. We construct a model assuming no dephasing and demonstrate a perfect match to our measurements indicating that the investigated quantum dot is indeed dephasing free. 
The discovery of dephasing free entanglement generation from a QD makes reaching perfect entanglement in the future merely a technical one.



We start by discussing the physics of how perfect entanglement between the biexciton and the exciton photon can be degraded. Due to the optical selection rules, the spin orientation of the decaying biexciton or exciton electron-hole pair projects onto a certain polarization state. Therefore, we must only analyze how the spins of the biexciton and exciton can loose their spin information.
For that, it is crucial to understand that the spin information, responsible for the entanglement generation, can only be destroyed through magnetic fields interacting with the exciton spin. The biexciton remains unaffected as its singlet state does not allow spin flips nor is its energy split through magnetic fields. Thus, we can solely focus our analysis on the exciton with its net one spin. The exciton is influenced by two kinds of magnetic fields. First, nearby spins carried by free or trapped \cite{PhysRevLett.105.187602} charge carriers or nuclei can lead to an interaction via spin flips. Second, effective magnetic fields caused by electric fields through the spin-orbit coupling can interact with the exciton spin. These electric fields can be decomposed into a static and a time varying contribution. Static fields can reduce the symmetry of the crystal field and are responsible for the so-called fine structure splitting \cite{PhysRevLett.103.063601} leading to a precession of the exciton spin \cite{WardNatComm, doi:10.1021/nl503581d}. Still, this effect is only unitary and leaves the entanglement of the state unaffected, and can be completely corrected for \cite{2017arXiv1707.04104, PhysRevA.73.033813}. However, fluctuating fields from free charge carriers and their spins can lead to dephasing of the quantum state. With a (quasi)-resonant excitation scheme spin and charge noise from free carriers can be greatly suppressed \cite{MullerNat2014}. Thus, under a (quasi)-resonant excitation scheme the magnetic field fluctuations from the nuclei should be the only remaining significant source of dephasing. In contrast to assumptions in other works \cite{Huber2017, Keil2017, PhysRevB.88.041306} we find that this is not a significant source of dephasing and reveal that these interactions are negligible.
\begin{figure}
 \centering
 \includegraphics[scale=1]{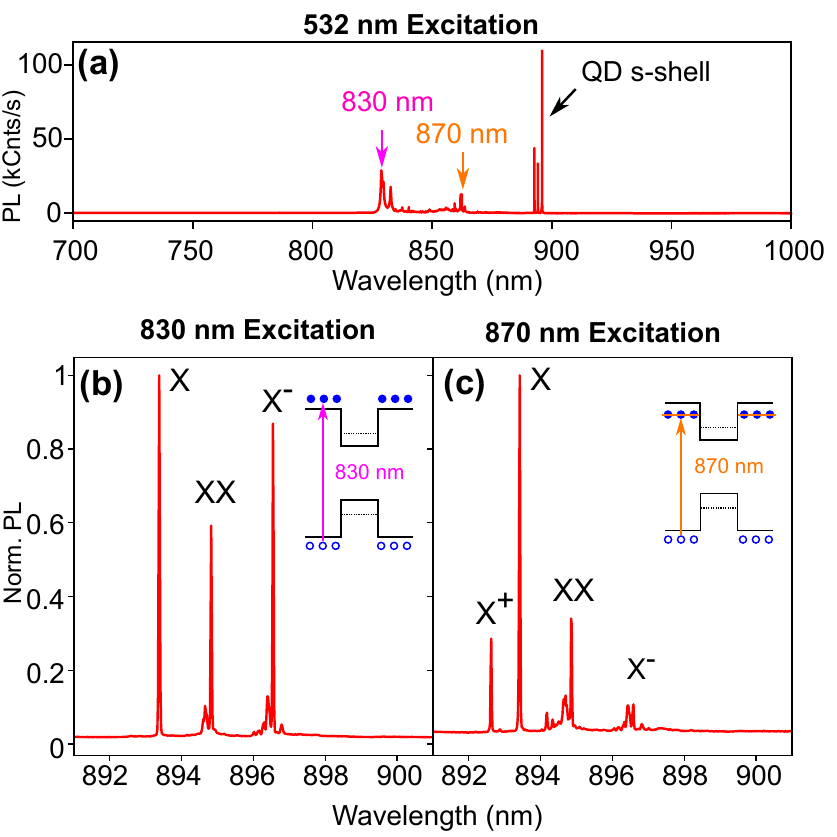}
 \caption{\textbf{QD emission spectra.} \textbf{(a)} Wideband emission spectrum excited with a green laser. For the entanglement measurements two excitation wavelengths have been used indicated by an arrow at $830\,\mathrm{nm}$ and at $870\,\mathrm{nm}$. \textbf{(b)} Higher resolution spectrum of the QD emission at $830\,\mathrm{nm}$ excitation showing three clean peaks attributed to the exciton ($X$), biexciton ($XX$), and negatively charged exciton ($\mathrm{X}^{-}$) at saturation. \textbf{(c)} Increasing the excitation wavelength to $870\,\mathrm{nm}$ leads to a different spectrum where $\mathrm{X}^{-}$ is suppressed and the positively charged exciton ($\mathrm{X}^{+}$) appears. All spectra in panels a-c were recorded at saturation. }
 \label{fig:qd}
\end{figure}

In previous work, it was shown that the nuclei field affects neutral excitons much less than charged exciton complexes \cite{RevModPhys.85.79}. Therefore, the neutral exciton should dephase on an even longer timescale than a charged exciton complex and be negligible during the radiative exciton decay of $\sim$~$1\,\mathrm{ns}$. This argumentation has been shown to be valid by a study revealing that the neutral exciton spin in InAs quantum dots is not affected by dephasing during the entire radiative lifetime of $\sim2.5\,\mathrm{ns}$\cite{Sénés2003}. Remarkably, this result was measured on a system with a large Indium content, an element with a significant nuclei spin of 9/2, which has been previously thought to limit dephasing free entanglement \cite{Huber2017, Keil2017, PhysRevB.88.041306}. Furthermore, spin-noise measurements \cite{Kuhlmann2013} suggest a strong noise suppression at frequencies on the timescale of the exciton lifetime.
Our measurements on a wurtzite InAsP quantum dot in a tapered InP nanowire \cite{doi:10.1021/nl303327h} are in perfect agreement with the above argumentation and reveal that under quasi-resonant excitation the exciton spin does not dephase over the entire exciton decay time of $\sim$~5\,ns. On the contrary, when excited non-resonantly the excess charges introduce significant dephasing setting in after $\sim$\,0.5 ns. 

Fig. \ref{fig:qd} (a) shows a photoluminescence (PL) spectrum of the QD under study indicating the resonances and the QD s-shell transitions. The peak at 830\,nm is the wurtzite InP nanowire band-gap transition \cite{doi:10.1021/nl303327h} and we excited the quantum dot at this wavelength to study the effect of dephasing. In contrast, for the dephasing free measurements we excited at $\sim870\,\mathrm{nm}$ where there are a manifold of peaks which stem from donor/acceptor excitons \cite{Skromme1983} and not from the QD's p-shell transitions since these lines were uncorrelated with the QD s-shell transitions. Due to the background n-doping ($\approx1\cdot10^{16}\,\mathrm{cm}^{-3}$, Supplementary Information of Ref. \cite{PhysRevB.93.195316}) of the nanowire the PL spectrum for the two excitation schemes is quite different. In the case of non-resonant excitation, shown in 
Fig. \ref{fig:qd} (b), only three clean peaks from the QD are visible attributed to the exciton ($X$), the biexciton ($XX$), and the negatively charged exciton ($X^-$). In case of quasi-resonant excitation ($830\,\mathrm{nm}$), Fig. \ref{fig:qd} (c), the $X^-$ is suppressed as the quantum dot s-shell is already filled with electrons due to the background n-doping and holes are more mobile so they can more readily relax into the quantum dot. Here, the positively charged exciton ($X^+$) dominates the $X^-$ line.

We now comment on the multiphoton emission of our entangled photon source, which degrades the entanglement fidelity, but is not a source of dephasing. Shown in Supplementary Information Fig. S1, the power dependent  $g^{(2)}$ remains flat at a level of $g^{(2)}(0)=0.003\pm0.003$ for the X and $g^{(2)}(0)=0.10\pm0.01$ for the $XX$ up until the $XX$ saturation point of $640\,\mathrm{nW}$. 
At XX saturation, detected count rates of $940\,\mathrm{kCnts/s}$ for the $X$ and $400\,\mathrm{kCnts/s}$ for the XX have been recorded with pulsed quasi-resonant excitation at a $76.2\,\mathrm{MHz}$ repetition rate and with the quantum state analysis optics removed. The setup efficiency in that case was $6.3\,\%$ from the first lens until a detected photon. This detected count rate corresponds to a high photon-pair source efficiency of $1.63\,\%$, which is two orders of magnitude brighter than a quantum dot entangled photon source in the bulk \cite{Joens2017}. 
\begin{figure*}
 \centering
 \includegraphics[scale=1]{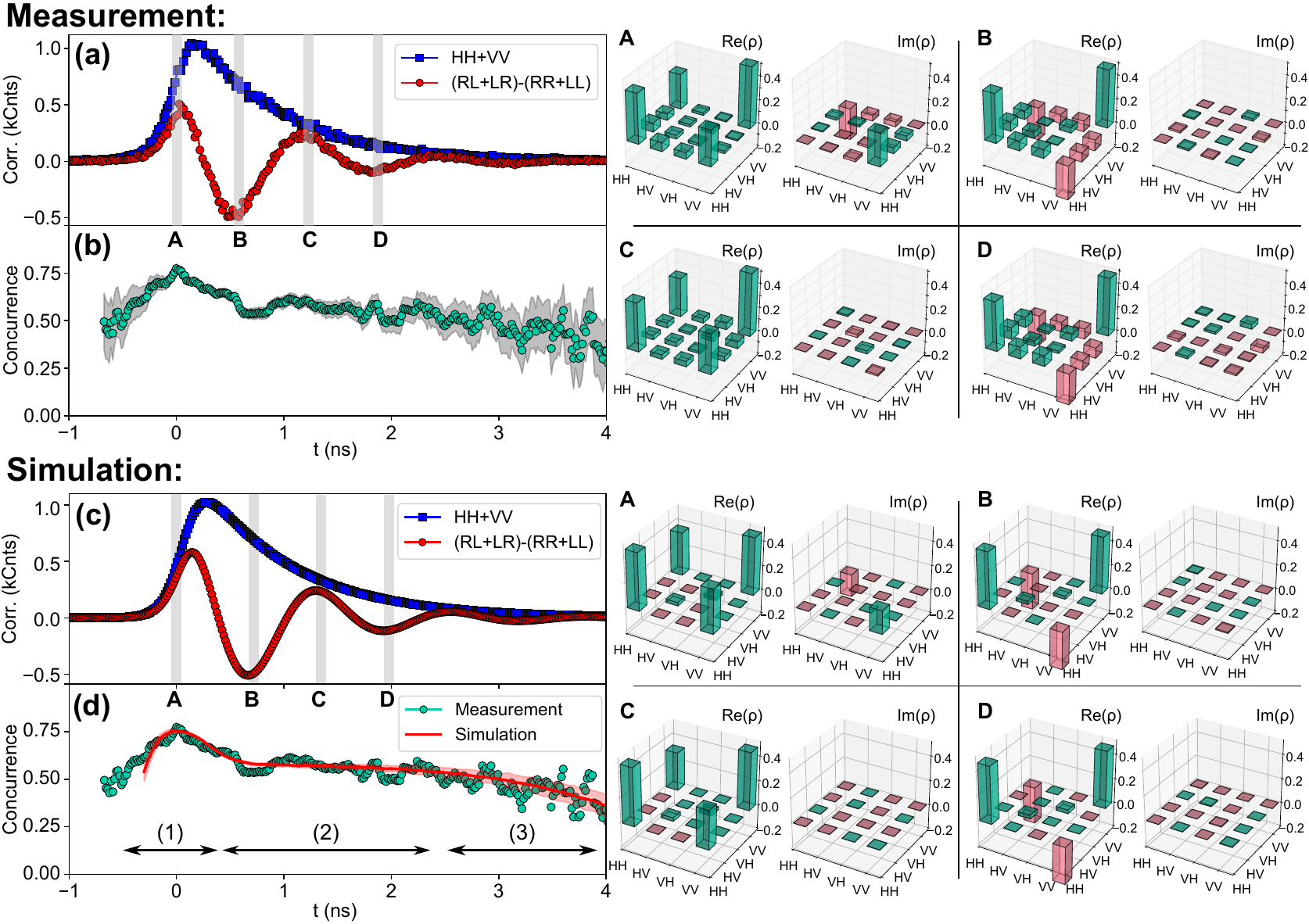}
 \caption{\textbf{Dephasing free entanglement.} \textbf{(a)} Two-photon correlation measurements depicting the sum of the $HH$ plus $VV$ projections together with $\left(RL+LR\right)-\left(RR+LL\right)$ showing quantum oscillations. The quantum oscillations appear because the latter term is proportional to  the difference of the Bell states $\Phi^{+}=1/\sqrt{2}\left(\ket{RL}+\ket{LR}\right)$ and $\Phi^{-}=1/\sqrt{2}\left(\ket{RR}+\ket{LL}\right)$. The gray shaded areas indicate times with the highest concurrence (\textbf{A}) and times with the smallest imaginary value of the density matrix (\textbf{B - D}). \textbf{(b)} The concurrence extracted from the measurement as a function of time delay, $t$, for all 36 bases. Each data point contains the correlation counts for a 100\,ps time window. The gray area indicates a $2\sigma$ concurrence error based on counting statistics. \textbf{(c)} The simulation shows the outcome of a fit free model of the quantum dot which is in perfect agreement with the measurement shown in \textbf{(a)}. The gray shaded areas indicate times with the highest concurrence (\textbf{A}) and times with the smallest imaginary value of the density matrix (\textbf{B - D}). \textbf{(d)} The concurrence measurement (green solid circles) is superimposed with the simulation (solid red line). The simulation is in excellent agreement with the measurement over the entire exciton lifetime indicating dephasing free entangled photon generation.}
  \label{fig:evolution}
\end{figure*}

In the following, we show that we have realized a dephasing free source of entangled photons. This remarkable finding implies that it is possible to reach perfect entanglement from quantum dots, which is in stark contrast to the common understanding \cite{Huber2017, Keil2017, PhysRevB.88.041306} that quantum dots cannot reach `perfect' entanglement due to dephasing mechanisms such as interaction with nuclei. To explain our findings we use a model of a dephasing free biexciton-exciton cascade.
First, we focus on the results of the quasi-resonant excitation scheme and find a perfect match to the dephasing free model. 
Second, we compare this quasi-resonant excitation scheme with non-resonant excitation at $830\,\mathrm{nm}$ to show the effect of dephasing.

\begin{figure*}
 \centering
 \includegraphics[scale=1]{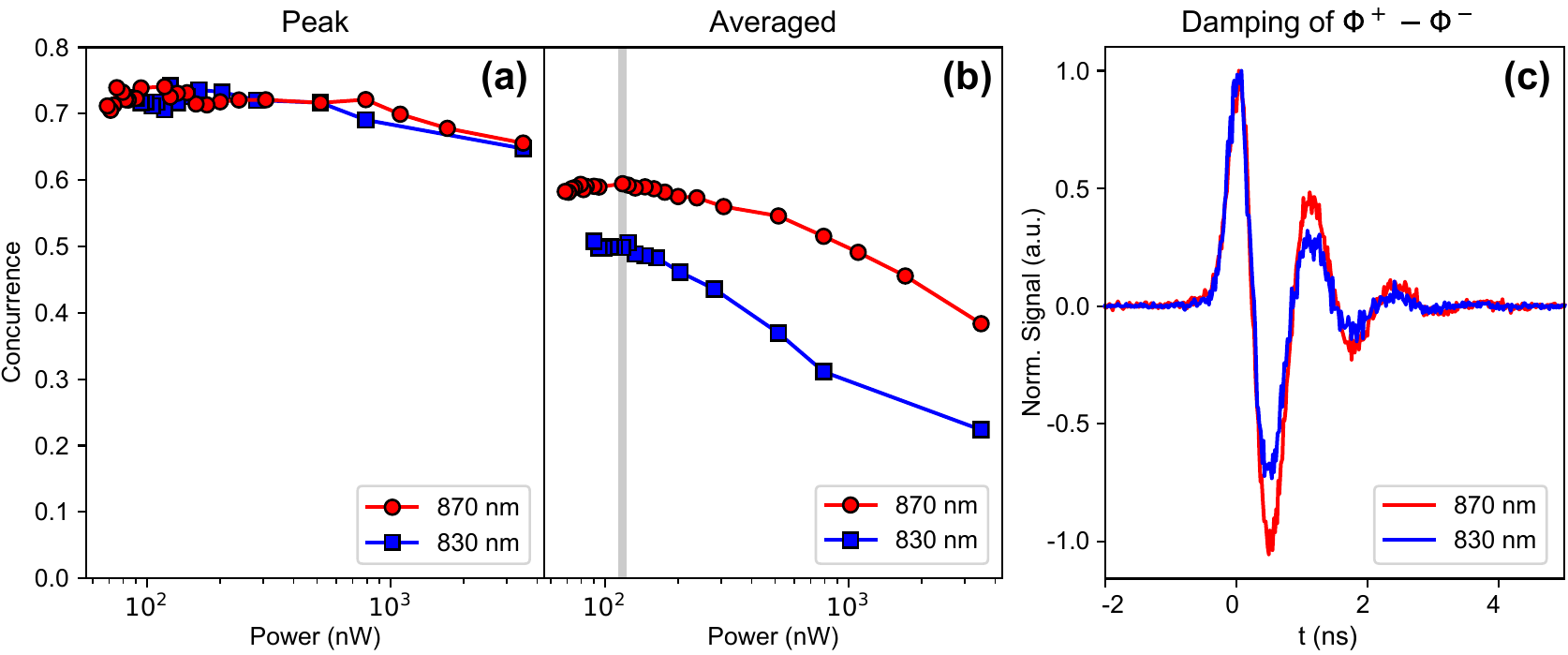}
 \caption{\textbf{Power dependent entanglement measurements.} \textbf{(a)} Peak concurrence calculated based on the two-photon correlation counts measured within a $200\,\mathrm{ps}$ wide window. There is no significant difference between the case of quasi-resonant excitation ($870\,\mathrm{nm}$) and non-resonant excitation ($830\,\mathrm{nm}$). \textbf{(b)} Counts averaged concurrence over the entire time window for $830\,\mathrm{nm}$ and 870\,nm excitation. In this situation, the non-resonant (830\,nm) is smaller than the quasi-resonant (870\,nm) excitation highlighting the effect of exciton dephasing. \textbf{(c)} The dephasing can also be visualized directly based on the normalized quantum oscillations when comparing both excitation schemes. The data is taken at the same excitation power as highlighted in gray from panel (b).}
  \label{fig:powerdep}
\end{figure*}


The entanglement results of the quasi-resonant excitation scheme are shown in Fig. \ref{fig:evolution} while the comparison between these two excitation schemes will be discussed subsequently. For the biexciton-exciton cascade we expect to measure a quantum state of the form \cite{PhysRevLett.101.170501}:
\begin{equation}
  \ket{\Psi(t,\delta)}=\frac{1}{\sqrt{2}}\left(\ket{HH}+e^{-i\frac{\delta}{\hbar}t}\ket{VV}\right)\Theta(t),
  \label{equ:state}
\end{equation}
where $\delta$ represents the fine structure splitting (FSS) energy, $t$ the time after the biexciton emission, and $\Theta(t)$ the Heaviside step function accounting for the fact that the X photon is created after the XX photon. 
We denote here the 36 possible correlations within a time interval $dt$ as $ij$ where $i,j\in \{H,V,D,A,R,L\}$ with the measurement polarization bases as $H/V$ (horizontal/vertical), $D/A$ (diagonal/anti-diagonal), and $R/L$ (right/left). Here, $i$ and $j$ represent polarization of the X and XX analyzer, respectively. With that, the likelihood $p_{ij}$ of measuring a correlation in the bases $ij$ within $dt$ reads as
\begin{equation}
  p_{ij}=\left({|\Braket{ij|\Psi(t)}|}^2 n(t,\tau_X)\right)*g(t)dt,
  \label{equ:projection}
\end{equation}
where $n(t,\tau_X)=1/\tau_X e^{-t/\tau_X}$ describes the probability of an exciton decay with time constant $\tau_X$, $*$ the convolution, and $g(t)$ denotes the detector systems' time resolution function. Therefore, the number of measured correlation counts per time bin becomes $ij=p_{ij}N_0 dt$ where $N_0$ is the number of collected biexciton-exciton pairs. 

Based on this mathematical description, the decay of the sum of the correlation counts $HH+VV$ is proportional to the exciton lifetime, $\tau_X$. We plotted the sum of these correlation counts, $HH+VV$, with blue squares in Fig.~\ref{fig:evolution} (a) from which we extracted $\tau_X=847\pm6\,\mathrm{ps}$. Furthermore, equation \ref{equ:state} describes an oscillation of the quantum state between the two Bell states $\ket{\Phi^{+}}=\frac{1}{\sqrt{2}}\left(\ket{RL}+\ket{LR} \right)$ and $\ket{\Phi^{-}}=\frac{1}{\sqrt{2}}\left(\ket{RR}+\ket{LL} \right)$ with a period of $\hbar/\delta$. Therefore, plotting the measured correlations $\left(RL+LR\right)-\left(RR+LL\right)$ reveals quantum oscillations \cite{WardNatComm,doi:10.1021/nl503581d} between the two Bell states as shown by red circles in Fig. \ref{fig:evolution} (a). The quantum oscillation allowed us to accurately measure the FSS to be $795.52\pm0.35\,\mathrm{MHz}$, an accuracy which is unachievable with typical spectroscopic techniques \cite{NanowireEntanglement}. We note that the exciton lifetime and FSS completely describe the quantum state evolution as noted in equation \ref{equ:state}.


For the entanglement measurements in Fig. \ref{fig:evolution} the QD was excited very close to saturation with an excitation power of $112\,\mathrm{nW}$. The correlations between the X and XX photons were measured in all possible 36 bases \cite{PhysRevA.78.052122} $ij$ instead of the minimal necessary \cite{PhysRevA.64.052312} 16. This enabled us to perform a better density matrix reconstruction based on a maximum likelihood approximation \cite{PhysRevA.64.052312, qtomo}. 
We calculated the density matrices using multiple time windows with a width of $dt=100\,\mathrm{ps}$ during the radiative decay of the exciton. Four representative density matrices are shown in the inset of Fig. \ref{fig:evolution}. Inset A represents the density matrix at the highest measured entanglement strength. Interestingly, there is an imaginary contribution even though equation \ref{equ:state} predicts no imaginary part at $t=0$. The cause of this effect is the finite time resolution of the employed avalanche photodiode single photon detectors that averages the phase of the exciton spin precession. In contrast, the density matrices presented in insets B, C, and D were chosen with the smallest imaginary parts. Similarly, the finite detector time resolution is responsible that smallest imaginary parts are not observed at the extrema of the quantum oscillations, but slightly time delayed.

For a complete picture of the entanglement time evolution the concurrence $\mathcal{C(\rho)}$, defined in Ref. \cite{PhysRevLett.80.2245}, is a more suitable way of analyzing the entanglement strength of the density matrix $\rho$. The concurrence scales between zero and one \cite{PhysRevLett.80.2245}, whereby it is one in the case of the system being fully entangled and zero if the system exhibits only classical correlations. Fig. \ref{fig:evolution} (b) shows the concurrence evolution as a function of time delay where each point was calculated based on the correlations within a $100\,\mathrm{ps}$ time window. The concurrence reaches a maximum of $\mathcal{C}=0.77\pm0.02$, while a counts weighted concurrence average over the whole time window yields $\bar{\mathcal{C}}=0.62\pm0.03$. 

In Fig. \ref{fig:evolution} (c-d) we compare the measured result with a simulation assuming a dephasing free QD without any free parameters. Our fit-free model, based on equation \ref{equ:projection}, only considers the finite detection time response, the dark counts, the FSS, the finite $g^{(2)}$ of the $XX$ photon, the detected count rates, and the exciton lifetime that were all determined from the experimentally measured ones. To get a more realistic implementation we added the detectors' dark counts before calculating the density matrix $\rho_{dc}$ based on a maximum likelihood approximation.
The finite $g^{(2)}_{XX}(0)$ of the biexciton will spoil the entanglement generation in $g^{(2)}_{XX}(0)$-fraction of the cases. Therefore, we can add-mix uncorrelated light to $\rho_{dc}$ as

\begin{equation}
  \rho_{sim}(t) = \left(1-g^{(2)}_{XX}(0)\right)\rho_{dc}(t) + \frac{g^{(2)}_{XX}(0)}{4}\mathbb{1},
\end{equation}
where $\rho_{sim}(t)$ is the result of the simulation, and $\mathbb{1}/4$ is the density matrix for uncorrelated light.
Remarkably, excellent agreement between the model and measurement is achieved without any free parameters. To get a more quantitative number it is best to compare the correlation counts weighted concurrence average $\bar{\mathrm{C}}$ over the full time window. From the simulation we obtain $\bar{\mathcal{C}}(\rho_{sim})=0.61\pm0.01$, whereas from the measurement this yields $\bar{\mathcal{C}}(\rho)=0.62\pm0.03$. These results agree within their error bounds, further exemplifying their perfect agreement. We therefore conclude that our quantum dot does not exhibit dephasing over its entire lifetime.

In addition to the excellent agreement of the concurrence, we also see that the density matrices match well between the measurement and the simulation as shown in the insets of A-D. Of particular interest is inset A. Both the simulation as well as the measurement exhibit non-vanishing imaginary parts. The reason for this observation is phase averaging during the exciton precession caused by the finite time resolution of the detectors. This effect has been seen before \cite{doi:10.1021/nl503581d}, but a convincing explanation has remained elusive.

Even more astonishing; however, is the perfect agreement of the concurrence simulation with the concurrence extracted from the measurement depicted in Fig. \ref{fig:evolution} (d). Here, we identify three regimes: (1) the `top'-part; (2) the `flat'-part; and (3) `roll-off'-part. The `top'-part exhibits a concurrence maximum because the concurrence first rises as the detector response function $g(t)$ samples more and more correlation counts with evolving time, $t$. At a certain level; however, the phase averaging of the exponential term in equation \ref{equ:state} dominates and the concurrence falls. Once the whole $g(t)$ function samples the state evolution, the phase averaging remains constant, named the `flat'-part. With evolving time less correlations are measured due to the exponential decay of the X photon, which is when we enter the `roll-off'-part where the concurrence drops due to the detector dark counts. It is important to note that the whole entanglement evolution with its three parts can be completely described without any dephasing from the QD. The three regimes are solely caused by the finite time resolution and dark counts from the detection system.  

\begin{figure*}
 \centering
 \includegraphics[scale=1]{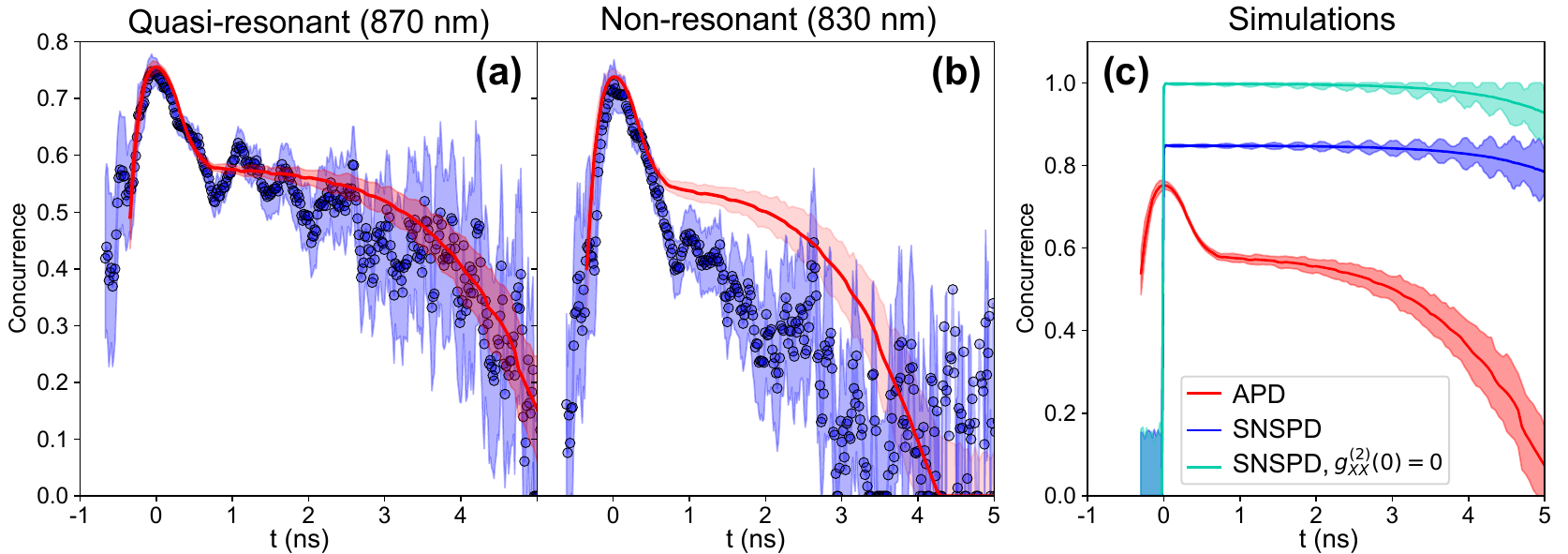}
 \caption{\textbf{Towards near-unity entanglement: comparison of dephasing and dephasing free entanglement.} \textbf{(a)} At quasi-resonant excitation the measured concurrence evolution agrees with the simulation within error bars, thus signifying dephasing free entanglement. \textbf{(b)} At non-resonant excitation the measured concurrence evolution does not match with the simulation indicating dephasing. The data in both (a) and (b) was taken from the two points highlighted in the gray region of Fig. 3b. \textbf{(c)} Three simulation curves illustrating the effect of finite detection time resolution and multiphoton emission of the biexciton photon. The red graph depicts the same simulation as already presented in Fig. \ref{fig:evolution} (d) with finite $g^{(2)}_{XX}=0.1$ and a slow detection system based on an avalanche photodiode single-photon detector (APD) as a reference. The blue curve shows the outcome of a simulation similar to the red curve with finite $g^{(2)}_{XX}=0.1$, but with a fast detection system based on a superconducting nanowire single photon detector (SNSPD) with $30\,\mathrm{ps}$ timing resolution. The cyan curve is the same as the blue curve with SNSPD, but for pure single photon emission of the biexciton photon ($\it{i.e.,}$ with $g^{(2)}_{XX}(0)=0$). Remarkably, with a fast detection system and perfect $g^{(2)}_{XX}$, near-unity entanglement is expected. Dark counts used in the simulation for the APDs are $36.3\,\mathrm{s^{-1}}$ and $18.2\,\mathrm{s^{-1}}$ for the X and XX detector respectively and for the SNSPDs are $1\,\mathrm{s^{-1}}$.}
  \label{fig:dephasing}
\end{figure*}

We now repeat the experiment with non-resonant excitation and compare it with the quasi-resonant excitation scheme. Fig. \ref{fig:powerdep}(a) depicts the peak concurrence for the two different excitation schemes as a function of excitation power. Each data point is constructed by analyzing the correlation counts within a $200\,\mathrm{ps}$ time window centered at $t=0$. The result reveals that both excitation schemes provide the same power dependent peak concurrence measurement. The cause for the concurrence to drop at higher excitation power is the increase of the biexciton $g^{(2)}$-value which is not a dephasing effect. The situation is quite different when we compare the correlation weighted concurrence average over the full time window as presented in Fig. \ref{fig:powerdep} (b). Clearly, the data for $830\,\mathrm{nm}$ excitation shows $\sim15\,\%$ smaller entanglement strength as compared to the quasi-resonant case at low powers, while deviating further at higher powers. This result is expected from the excess charges generated by non-resonant excitation. First, the spin of these charges can cause direct spin flip-flop processes with the exciton spin. Second, fluctuating electric fields caused by the excess charges can result in an effective magnetic field via the spin-orbit interaction and alter the exciton spin. This situation is directly visible in Fig. \ref{fig:powerdep} (c) where the normalized quantum oscillations are compared with each other. The two curves were recorded at the same power level highlighted by the shaded region in Fig. \ref{fig:powerdep} (b). The $830\,\mathrm{nm}$ data clearly damps out faster than the $870\,\mathrm{nm}$ one, which is the fingerprint for dephasing.

To show the dephasing effect more quantitatively we simulated the two cases as presented in Fig. \ref{fig:dephasing}. 
Again, the model in Fig. \ref{fig:dephasing} (a) agrees with the quasi-resonant excitation scheme within error bars indicative for dephasing free entanglement. In contrast, the situation is very different for non-resonant excitation (Fig. \ref{fig:dephasing} (b)) where the simulation clearly overestimates the measurement exemplifying dephasing.

Finally, we investigate how the concurrence evolution of the data presented in Fig. \ref{fig:evolution} would look like if we were to measure with an emerging detection system employing a better timing resolution and lower dark count rate. We assume a time resolution of $30\,\mathrm{ps}$ full width at half maximum and a dark count rate of $1\,\mathrm{Hz}$, values which can be met by recently available superconducting nanowire single photon detectors. The outcome of such a simulation is plotted in Fig. \ref{fig:dephasing} (c) for $g^{(2)}_{XX}(0)=0$ and  $g^{(2)}_{XX}(0)=0.1$ and is compared to the case when measuring with our APDs and  $g^{(2)}_{XX}(0)=0.1$ ($\it{i.e.}$, with the same plot as in Fig. \ref{fig:evolution} (d)). In both cases with $30\,\mathrm{ps}$ timing resolution, $\it{i.e.}$, with zero $g^{(2)}_{XX}(0)$ and finite $g^{(2)}_{XX}(0)$, the difference to the simulation with APDs is quite striking. First, the peak concurrence for finite $g^{(2)}_{XX}(0)=0.1$ ($C=0.849\pm0.001$) and  $g^{(2)}_{XX}(0)=0$ ($C=0.999\pm0.001$) is significantly larger than for the case of measuring with APDs ($C=0.75\pm0.01$). Remarkably, the concurrence reaches near-unity for zero $g^{(2)}(0)$. Second, the `top'-part is completely suppressed. Instead, only the `flat'- and `roll-off'-parts remain. It is interesting to note that even the small dark count rate of $1\,\mathrm{Hz}$ is inducing a resolvable entanglement roll-off. However, this decrease at the end has minimal effect to the overall concurrence and a count averaged concurrence of $\bar{\mathcal{C}}=0.996^{+0.004}_{-0.008}$ for zero $g^{(2)}_{XX}(0)$ and $\bar{\mathcal{C}}=0.847\pm0.007$ in case of finite $g^{(2)}_{XX}(0)$ is obtained. The slight oscillations visible in the concurrence's error for the high temporal resolution simulation are not artefacts. They are caused by counting statistics since every time one of the 36 simulated correlations reaches zero the concurrence can be less accurately estimated. This happens with a frequency four times larger then the FSS. In fact, this effect is visible in other groups' measurements, for example in the fidelity evolution of Ref. \cite{WardNatComm}. For a slower detection system, such as our APDs, this effect is averaged out.

We have shown that our model is capable of explaining our measurement results in great detail. The question arises why dephasing free entanglement from QDs has not been observed before. The reason is that a QD with a long $X$ decay time of $\tau_X\sim1\,\mathrm{ns}$ is needed in conjunction with a (quasi)-resonant excitation scheme. For example, in Refs. \cite{MullerNat2014, Huber2017} a resonant excitation scheme was employed, but the QDs had a $\tau_X\sim200\,\mathrm{ps}$ which makes it very difficult to separate the detrimental effects from the detection system. However, based on model calculations, we predict that the QDs investigated in these aforementioned works  of Refs. \cite{MullerNat2014, Huber2017} should also be dephasing free. 
Therefore, the occurrence of dephasing free entanglement is not at all limited to InAsP QDs, but should be equally achievable in other QD materials such as InGaAs \cite{MullerNat2014}, and GaAs \cite{Huber2017}.

Even though we have found dephasing free entanglement, we have not yet shown unity entanglement. The reduction of the measured entanglement from unity comes mainly from the detectors' time resolution, finite $g_{XX}^{(2)}$-value of the QD, and dark counts. Still, $g^{(2)}$-values of both the exciton and biexciton can be brought to zero by resonant excitation \cite{MullerNat2014, Huber2017, SomaschiNat}. Therefore, the problem of reaching perfect entanglement from QDs should now be merely a technical one in future work by combining the right source and excitation scheme with a state-of-the-art detection system.

In conclusion, we have established a precise model of the entanglement measurement in which the generation and the detection processes of entangled photon pairs are of equal importance.
Based on this knowledge we could show that a QD containing Indium generates dephasing free photon entanglement over the entire exciton decay time even though its large nuclear spin of 9/2. 
This result is remarkable as it was thought to be unachievable due to interaction with the nuclei.
The conditions to be able to resolve dephasing free entanglement are (quasi)-resonant excitation and a precise knowledge of the employed detection system. This new insight will allow to make an ideal entangled photon source based on QDs.
We predict with our model that dephasing free entanglement is also found in materials other than InAsP, such as InGaAs \cite{MullerNat2014} and GaAs \cite{Huber2017} QDs.

A. Fognini gratefully acknowledges the Swiss National Science Foundation for the support through their Early PostDoc Mobility Program. K.D. J\"ons acknowledges funding from the Marie Sk\l{}odowska Individual Fellowship under REA grant agreement No. 661416 (SiPhoN). M.E. Reimer acknowledges Industry Canada and NSERC for support.

\section{Methods}
\subsection{Quantum Dot}
The quantum dot growth is described in the Methods section of Ref. \cite{NanowireEntanglement}.

\subsection{Measurement}
We used a standard micro-PL setup where the nanowire sample was kept at a base temperature of $4.5\,\mathrm{K}$. The light from a picosecond pulsed laser was filtered with a $1200\,\mathrm{lines/mm}$ grating to reduce the effect of laser background fluorescence before it was used to excite the QD. 
For the quantum state tomography we used a similar system as in Ref. \cite{Joens2017} with the difference that the waveplates where mounted in high-precision motorized mounts crucial for the repeatability of the experiment. The first beam splitter used to excite the QD had $30\,\%$ reflection and $70\,\%$ transmission. The excitation was performed in all cases with s-polarized light to prevent nuclear polarization \cite{PhysRevB.78.205325}.
All correlation data was sampled with $16\,\mathrm{ps}$ resolution. 

The data presented in Fig. \ref{fig:evolution} was excited with a power of $112\,\mathrm{nW}$ and integrated for $370\,\mathrm{s}$ per basis. Count rates were in HH basis $71\,\mathrm{kCnt/s}$ for the X and $8\,\mathrm{kCnts/s}$ for the XX.
The data presented in Fig. \ref{fig:dephasing} (a)/(b) was excited with a power of $118\,\mathrm{nW}$ and integrated for $342\,\mathrm{s}$ per basis. In case of Fig. \ref{fig:dephasing} (a) this resulted in a HH basis count rate of $85\,\mathrm{kCnt/s}$ for the X and $11\,\mathrm{kCnts/s}$ for the XX and for (b) in a HH basis count rate of $73\,\mathrm{kCnt/s}$ for the X and $4.4\,\mathrm{kCnts/s}$ for the XX.


The employed avalanche single-photon detectors (APDs) had a dark count rate of $36.3\,\mathrm{s^{-1}}$ for the detector measuring the exciton and $18.2\,\mathrm{s^{-1}}$ for the biexciton detector with a time resolution of $190\,\mathrm{ps}$ full width at half maximum. 

\subsection{Simulation}
For the simulation in the text we used a $\mathrm{FSS}=795.520\,\mathrm{MHz}$, a dark count rate of $36.3\,\mathrm{s^{-1}}$ for the exciton and $18.2\,\mathrm{s^{-1}}$ of the biexciton detector, a exciton lifetime of $\tau_X=847\,\mathrm{ps}$, a $g_{XX}^{(2)}=0.1$, $g_{X}^{(2)}=0$, and a laser repetition rate of $76.2\,\mathrm{MHz}$. The used count rates and integration times are stated in the Measurement section. In case of Fig. \ref{fig:evolution} (b) an exciton lifetime of $\tau_X=753\,\mathrm{ps}$ was used.

The density matrix reconstruction was performed based on the code from Ref. \cite{qtomo}. 
The method of how to acquire the system's time resolution $g(t)$ is described in the Supplementary Information.

The error of the concurrence is estimated based on a Monte-Carlo simulation assuming counting statistics. For each concurrence value the simulation was performed with 1000 repetitions.

\subsection{Author Contribution}
A.F., A.A., and M.Z. performed the experiments. A.F. developed and performed the data analysis with A.A..  
The sample was characterized by J.T.F..
A.A., M.Z., S.J.G., N.S., S.J.D, and M.E.R. built the setup. S.J.G., N.S., S.J.D., K.D.J., and M.E.R. helped during the experiment.
D.D. and P.J.P. grew the nanowire quantum dot sample.
A.F. and M.E.R. took the lead in writing the paper. V.Z. and M.E.R supervised the project.

%
\end{document}